\documentclass[12pt,twoside]{article}
\usepackage{graphicx}
\usepackage{xcolor}
\usepackage{amssymb}
\usepackage{amsmath}
\usepackage{amsfonts}

\let\UnmodifSec=\section
\renewcommand{\section}{\setcounter{equation}{0}\UnmodifSec}

\newtheorem{lemma}{Lemma}[section]

\def\bC{{\bf C}}
\def\bR{{\bf R}}

\def\bZ{{\bf Z}}
\def\Im{\mathop{\rm Im}\nolimits}
\def\Re{\mathop{\rm Re}\nolimits}
\def\Arg{\mathop{\rm Arg}\nolimits}

\def\th{\mathop{\rm th}\nolimits}
\def\ch{\mathop{\rm ch}\nolimits}

\def\Arctg{\mathop{\rm Arctg}\nolimits}

\def\FF{{\cal F}}

\def\ovl{\overline}

\def \vhi{\varphi}
\def \veps{\varepsilon}

\def\haf{{1 \over 2}}
\def\interior#1{\setbox1=\hbox{$#1$}\rlap{$#1$}\kern0.4\wd1\raise1.1\ht1%
\hbox{$\scriptstyle \circ$}}

\def\boxit#1#2{\setbox1=\hbox{\kern#1{#2}\kern#1}%
\dimen1=\ht1 \advance \dimen1 by #1 \dimen2=\dp1 \advance \dimen2 by #1
\setbox1=\hbox{\vrule height\dimen1 depth\dimen2\box1\vrule}%
\setbox1=\vbox{\hrule\box1\hrule}%
\advance \dimen1 by .4pt \ht1=\dimen1 \advance \dimen2 by .4pt \dp1=\dimen2
\box1\relax}
\def\endprf{\raise .5ex\hbox{\boxit{2pt}{\ }}}

\def\ifundefined#1{\expandafter\ifx\csname#1\endcsname\relax}

\def\beq{\begin{equation}}
\def\endq{\end{equation}}
\def\beqa{\begin{eqnarray}}
\def\endqa{\end{eqnarray}}

\def\k{\kappa}

\def\dateline{May 28, 2019}

\date{\dateline}
\title{A rigorous lower bound on the scattering
amplitude at large angle}
\author{Henri Epstein$^{(a)}$ and Andr\'e Martin$^{(b)}$\\
  $^{(a)}$IHES, Bures-sur-Yvette, France\\
  $^{(b)}$ CERN, Geneva, Switzerland}
\begin{document}
\parindent = 0pt
\maketitle
\vskip 3mm
\begin{abstract}
We prove a lower bound for the modulus of the amplitude for
a two-body process at large scattering angle. This is based on the interplay
of the analyticity of the amplitude and the positivity properties of
its absorptive part. The assumptions are minimal, namely those of
local quantum field theory (in the case when dispersion relations hold).
In Appendix A, lower bounds for the forward particle-particle and
particle-antiparticle amplitudes are obtained. This is of independent interest.
\end{abstract}

\vskip 1cm

\section{Introduction}
In 1963, F. C\'erulus and one of us (A.M.) obtained a lower bound on the
scattering amplitude at large angles \cite{cermartin64}.
It is not exactly a lower bound
at a given angle, because it is impossible to prevent the scattering amplitude
from vanishing at a given point. What we obtained is a lower bound on the
{\bf maximum of the modulus of the amplitude}  in some finite angular interval.
The assumption made was the validity of the Mandelstam representation with a
{\bf finite number of subtractions} \cite{mandelstam}.
Another assumption was that the
forward scattering amplitude cannot decrease faster than a power of $s$,
the square of the C.M. energy, but this assumption could be removed by
Jin and Martin in 1964 \cite{jinmartin64}, because at least for $\pi\pi$,
$p\pi$, $K\pi$ scattering
the forward amplitude cannot decrease faster than $1/s^2$, because of
the positivity of the absorptive part in the forward direction.
More exactly, there is at least a sequence of energies going to infinity
for which this is true. The lower bound was
$$
C \exp(-\sqrt{s}\ln s)
$$
where $s$ is the square of the center of mass energy.
At the time, it was a rather good surprise because experiments done by
the Cornell group indicated  that the large angle proton-proton
scattering amplitude
behaved like $C\exp(-\sqrt{s})$ \cite{baker6264}.
So we disagreed with experiment
only by the factor $\ln s$ in the exponential.

\vskip 3mm
It happens that this lower bound is violated by the Veneziano amplitude
\cite{veneziano68}. This is not astonishing, because the Veneziano amplitude implies
linear rising Regge trajectories  which implies an infinite number of
subtractions. So G.~Veneziano asked us if there exists a "rigorous lower bound"
(in the sense previously given) on the large angle scattering amplitude.
After the work of one of us on the enlargement of the domain of analyticity
in local field theory by using positivity properties of the absorptive part
\cite{martin66} and also the obtention of a lower bound of the forward
absorptive part \cite{cornille71} this turned out to be
possible, but the problem was left unsolved for about 50 years.
In the present work we give an answer, probably the best possible one,
but this answer is unfortunately an extremely small lower bound, so that
we publish our results rather as a matter of principle.

\vskip 3mm
The strategy is the following:
\vskip 3mm

1) from \cite{martin66} the absorptive part of the scattering amplitude is
analytic in an ellipse in the $\cos \theta$ variable with foci at $\pm 1$
and right extremity at     
$\cos \theta= 1+ 2 m_\pi^2/ k^2$, $k$ being the c.m. momentum.
The modulus of the absorptive part is maximum at the right extremity of the
ellipse, and {\bf morally} the absorptive part is bounded  by
$s^2$ in the ellipse.
{\bf Morally} means that we only know
from \cite{martin66} that the integral of the absorptive
part, divided by $s^3$, over $s$, is convergent for fixed momentum transfer,
$t = 4m_\pi^2 - \veps$, $\veps > 0$ arbitrarily small. 
In the special case of $\pi\pi$ scattering this integral is completely under
control
from tha absolute bounds obtained previously
(\cite{martin6465}, \cite{lukamar67}, \cite{bolomenn75}).

\vskip 3mm
2) as we said before, the forward scattering amplitude has a {\bf moral}
lower bound $1/s^2$,
and, since the diffraction peak cannot be arbitrarily small because
of the size of the ellipse, the elastic cross.section has a lower bound and
hence the absorptive part has a lower bound which
is $1/ s^5(\log s)^2$ \cite{cornille71}.

\vskip 3mm
      
3) If we have an upper bound on some angular interval of 
$-a<\cos \theta<+a$ and
also an upper bound on the border of the ellipse we can interpolate between
these 2 bounds 
because the logarithm of the modulus of the absorptive part
is a subharmonic function (of $\cos \theta$),
and we can get an upper bound anywhere inside the ellipse, in
particular at $\cos \theta= 1$, i.e. the forward direction. If the 
postulated bound on  the interval $(-a,\ +a)$
is too low we will get a contradiction
with the lower bound that we have in the forward direction. So the postulated
upper bound on $(-a,\ +a)$ cannot be arbitrarily small.

\vskip 3mm

However, things are not as simple as that because the lower bound on the
scattering amplitude is only  for discrete values of the energy,
and also, even if you assume that everything is continuous, you do not know if,
precisely, for these discrete values, the absorptive part, in the ellipse is
bounded by $s^2$. To overcome this problem we replace the scattering amplitude
by an average over some energy interval. The average
has all the nice properties we want, but, as we shall see, we loose 1 power
of s in the bound on the absorptive part, but this is unimportant.
The next section is devoted to this averaging. The following section explains
the interpolation described in 3) and gives the results. Details are given in
the appendices.

\section{Necessity of an averaging of the scattering amplitude on the energy}
  
1) As stressed by Common \cite{common70} and Yndurain \cite{ynd70},
we only know from \cite{martin66} that the
integral over the absorptive part $A_s$ :
\beq
\int_{(M_1+M_2)^2}^\infty {A_s(s,\ t= 4m_\pi^2 - \veps)\over s^3}\,ds
\endq
is convergent. This does not mean that $A_s$ is less than $s^2$. In fact 
$A_s$ can be very large or even infinite for isolated values of $s$.

\vskip 3mm
2) Since the forward scattering has a positive imaginary part for $s>0$
and a negative imaginary part for $s<0$ it  has been shown (\cite{jinmartin64}),
for a crossing-symmetric amplitude, that 
$\limsup s^2|F(s,0)| >0$. This means that there is a sequence of values
of $s$, $\{s_i\}$, going to infinity,
for which $s_i^2|F(s_i,0)>0$. In fact, in all 
non-real directions, $\lim s^2|F(s,0)|> 0$.
For non crossing-symmetric amplitudes, such as
$\pi^+ p \rightarrow \pi^+ p$ and $\pi^- p \rightarrow \pi^- p$,
we show in the present paper that
$\lim s^{2+\epsilon}|F(s,0)|> 0$, $\epsilon >0$ arbitrarily small,
{\it both} for $s \rightarrow +\infty$ and $s \rightarrow -\infty$
(i.e. $u \rightarrow +\infty$). This will be proved in Appendix \ref{lowerbd}.
So our results hold also for a non symmetric amplitude. The presence of
$\epsilon$ is inessential.

\vskip 3mm
3) Concerning the absorptive part, we have, from the optical theorem,  
$A_s(s,0)> s \sigma_{\rm total}> s \sigma_{\rm elastic}$,
and we need a lower bound on $\sigma_{\rm elastic}$.
Assume {\bf provisionally} that, for $t= 4m_\pi^2-\veps$,
the absorptive part is bounded by $s^N$.
We know that we cannot really assume that, but it will be corrected later
by averaging. Then, using Schwarz's inequality, we have
\begin{align}
\left | \sum_{l=0}^\infty (2l+1) f_l \right | &\le
\left| \sum_{l=0}^L (2l+1) f_l \right |+
\left| \sum_{l=L+1}^\infty (2l+1) f_l \right | \cr
&\le \sqrt{ \left ( \sum_{l=0}^L (2l+1) |f_l|^2 \right )}\times (L+1)
+\left| \sum_{l=L+1}^\infty (2l+1) f_l \right | \cr
&\le \sqrt{ \left ( \sum_{l=0}^L (2l+1) \Im f_l\right )}\times (L+1)
+\left| \sum_{l=L+1}^\infty (2l+1) f_l \right | \ ,
\label{n.10}\end{align}
where $L$ will be chosen later. If the absorptive part is bounded
by $s^N$ for $T < 4m_\pi^2$,
\beq
\left |\sum (2l+1)\Im f_l P_l\left ( 1+{T\over 2k^2} \right ) \right | < s^N\ ,
\label{n.15}\endq
so by Schwarz's inequality
\beq
\left | \sum_{L+1}^\infty (2l+1) f_l \right |^2 \le
\sum_{L+1}^\infty {2l+1\over P_l\left ( 1+{T\over 2k^2} \right )}
\sum_0^\infty (2l+1) P_l \Im f_l\ .
\label{n.20}\endq
The second factor in (\ref{n.20}) is bounded by $s^N$. The first factor
can be calculated by using the inequality
\beq
P_l(x) > {1\over 3} \left ( x+ \haf\sqrt{x^2-1} \right )^l\ \ {\rm for}\ x> 1
\label{n.21}\endq
which follows from the integral representation of Legendre polynomials.
We get, choosing $L=P\sqrt{s}\log s$, an asymptotic upper bound
\beq
s^{-P{\sqrt t\over 2\sqrt{2}}} \times {3P s\log s\over \sqrt{t}}\ .
\label{n.25}\endq
Choosing $P$ large enough compared to $N$, the second term in (\ref{n.10})
is negligible and we get
\beq
|F(s,\ 0)|^2 < P^2 s (\log s)^2 A_s(s,\ 0)\ .
\label{n.30}\endq
If
\beq
|F(s,\ 0)| > {1\over s^{2 \pm \veps}}\ ,
\label{n.35}\endq
we get
\beq
A_s > C {1\over s^{5 \pm 2\veps}(\log s)^2}\ .
\label{n.36}\endq
Unfortunately we cannot assume $A_s(s,\ T) < s^N$, but this will be remedied
now by averaging the amplitude over an energy interval.

\vskip 3mm
To solve the problem we propose to make an average over energies.
We define
\begin{align}
f(s,\ \cos \theta) &= F(s,\ t,\ u),\cr
\cos \theta &= 1 + {t\over 2k^2}\ .
\label{n.40}\end{align}
Now we average $f$ over an energy interval $\Delta$ :
\beq
f_\Delta(s,\ \cos \theta) = {1\over \Delta}
\int_{s-\Delta}^s f(s',\ \cos \theta)\,ds'\ .
\label{n.41}\endq
This averaging is mainly interesting for $s$ physical. However in the special
case of $\cos \theta =1$, it remains meaningful for $s <0$, i.e.
for the $u$ channel. We take $\Delta$ smaller than the interval
between the left cut and the right cut, and it follows that
\beq
\limsup_{s \rightarrow +\infty}|f_\Delta(s,\ 1)| s^{2+\veps} > 0
\label{n.42}\endq
as in the case of $f$.

The reason why we take this averaging is that
unitarity of the partial waves survives :
\beq
f_{\Delta l}(s) = {1\over \Delta} \int_{s-\Delta}^s f_l(s')\,ds'\ .
\label{n.45}\endq
Now
\beq
\Im f_{\Delta l}(s) = {1\over \Delta} \int_{s-\Delta}^s \Im f_l(s')\,ds'
> {1\over \Delta} \int_{s-\Delta}^s |f_l(s')|^2\,ds'\ .
\label{n.46}\endq
But by Schwarz's inequality
\beq
{1\over \Delta} \int_{s-\Delta}^s |f_l(s')|^2\,ds' 
\ge {1\over \Delta^2}\left |\int_{s-\Delta}^s |f_l(s')|\,ds'\right |^2\ ,
\label{n.47}\endq
so
\beq
\Im f_{\Delta l}(s) > |f_{\Delta l}(s)|^2\ .
\label{n.48}\endq
The absorptive part
\beq
a_\Delta(s,\ \cos \theta) =
\sum_{l=0}^\infty \Im f_{\Delta l}(s) P_l(\cos \theta)\ ,
\label{n.50}\endq
\beq
a_\Delta(s,\ \cos \theta) = {1\over \Delta} \int_{s-\Delta}^s
A_s \left (s',\ t= (\cos \theta -1) 2{k'}^2 \right )\,ds' \ .
\label{n.51}\endq
For $\cos \theta >1$ we see that in the integrand
$A_s \left (s',\ t= (\cos \theta -1) 2{k'}^2 \right )$ is less than
$A_s \left (s',\ t= (\cos \theta -1) 2{k}^2 \right )$ since,
in the interval $0 \le t < 4m_\pi^2$, $A_s$ increases
since the Legendre polynomial expansion converges.

Now remember that, according to \cite{martin66},
\beq
\int_{(M_A+M_B)^2}^\infty {A_s(s',\ t)\over {s'}^3} ds'\ \ \ 
{\rm converges\ for}\ \ t< 4m_\pi^2\ .
\label{n.55}\endq

Hence, for fixed $s \ge (M_A+M_B)^2 +\Delta$ and $t< 4m_\pi^2$,
\begin{align}
a_\Delta \left (s,\ \cos \theta = 1+ {t\over 2k^2} \right ) &\le {1\over \Delta}
\int_{s-\Delta}^s A_s \left (s',\ t\right )\,ds'\cr
&\le {s^3\over \Delta}\int_{(M_A+M_B)^2}^\infty
{A_s\left (s',\ t\right )\over {s'}^3} ds'\ .
\label{n.55.1}\end{align}

We conclude that
\beq
a_\Delta \left (s,\ \cos \theta = 1+ {t\over 2k^2} \right ) < C s^3
\ \ \ {\rm for}\ \ t< 4m_\pi^2\ .
\label{n.60}\endq
So the previous argument, applied to $A_s$ with the assumption
$A(s,\ t) < s^N$ for $t< 4m_\pi^2$, applies also to $a_\Delta$ with $N=3$.
We realize that we are loosing one power of $s$, but this is unimportant.
We conclude that
\beq
a_\Delta(s_i,\ \cos \theta =1) > {C\over s_i^{5+2\veps}(\log s_i)^2}
\label{n.65}\endq
where $s_i$ belongs to the sequence where
$|f_\Delta(s,\ \cos \theta = 1)| s^{2+\veps}$ approaches infinity.

\section{The least upper bound for $|\cos \theta| \le a$}
Now we have all the ingredients to find a least upper bound of the
scattering amplitude in an angular interval $|\cos \theta| \le a$, for instance
$|\cos \theta| \le \haf$ :
\beq
{\rm 1)}\ \ a_\Delta(s_i,\ \cos \theta =1) > {C\over s_i^{5+2\veps}(\log s_i)^2}\ ,
\label{n.70}\endq
\beq
{\rm 2)}\ \ a_\Delta \left (s,\ \cos \theta =
{4m_\pi^2 -\eta\over 2k^2} \right ) < Cs^3\ .
\label{n.71}\endq
This bound, because of the positivity of the $\Im f_{\Delta l}$'s
holds in the whole ellipse with foci at $\cos \theta = \pm 1$
and extremity at $\cos \theta = 1+{4m_\pi^2 -\eta\over 2k^2}$,
$\eta$ positive arbitrarily small.

\vskip 3mm
To interpolate between the bounds (\ref{n.70}) and (\ref{n.71})
we use the following
fact (proved in Appendix \ref{interp}) : let $f$ be a function holomorphic
in the domain $D_L$ bounded by the ellipse
$E_L$ with foci $\pm 1$ and semi-great axis $\ch(L)$, $L>0$,
\begin{align}
E_L &= \{ z\ :\ z = \cos(\theta + iL),\ \theta \in \bR\}\ ,\cr
D_L &= \{ z\ :\ z = \cos(\theta + iy),\ \theta \in \bR,\ \ 
y \in \bR,\ \ |y| < L\}\ .
\label{n.100}\end{align}
We suppose that $|f| \le M$ on $D_L$
and $|f(z)| \le m$ $\forall z \in [-a,\ a]$, where $0<m<M$
and $a=\cos(b)$, $0< b< \pi/2$.
Then
\beq
|f(1)| < M^{1-\alpha}\,m^\alpha\ ,
\label{g.10}\endq
where, for very small $L$,
\beq
\alpha = {4\over\pi} \exp\left({-\pi b\over 2L}\right )
\label{g.11}\endq
($\alpha$ is the quantity denoted $1-H(1)$ in Appendix \ref{interp}).
We can rewrite (\ref{g.10}) in the form
\beq
m \ge M \left ( {|f(1)|\over M}\right ) ^{1\over \alpha}\ .
\label{g.12}\endq

\vskip 0.3cm
We apply this to the case when 
$z = \cos(\theta)$, $f(z) = a_\Delta(s,\ \cos(\theta))$,
and $s$ belongs to a certain real sequence tending to infinity
such that (\ref{n.70}) holds. We suppose, for simplicity, that
all masses are equal to 1.
We choose $a = \haf \Rightarrow b = \pi/3$.
The half great axis of the ellipse is $\ch(L)$,
\beq
\ch(L) = 1+ {4\over 2k^2},\ \ k^2 = {s\over 4}-1\ ,\ \
{L^2\over 2} \sim {8\over s},\ \ L\sim {4\over \sqrt{s}}\ .
\label{g.20}\endq
Hence
\beq
\alpha = {4\over\pi} \exp\left({-\pi^2 \sqrt{s}\over 24}\right ).
\label{g.21}\endq
According to (\ref{n.71}), we can take $M= C_1s^3$. 
On the other hand, by (\ref{n.70}),
\beq
|f(1)| = |a_\Delta(s,\ \cos(\theta) = 1)| > {C_2\over s^5 \log(s)^2}\ .
\label{g.30}\endq
Therefore (using (\ref{g.12}))
\beq
m = \sup_{|\cos(\theta)| < \haf}|a_\Delta(s,\ \cos(\theta)| 
\ge C_3 s^3
\Big [ C_4 s^8 \log(s)^2 \Big ]^{-{\pi\over 4}
\exp\left({\pi^2 \sqrt{s}\over 24}\right )}\ .
\label{g.40}\endq
This clearly implies similar lower bounds for $A_s(s,\ t)$ and $F(s,\ t)$.

This result is rather disappointing but, as a matter of principle,
we see that {\it it is not zero}. For different values of $a$
we get the same qualitative behavior. For a non-symmetric amplitude,
like $\pi^+\pi^- \rightarrow \pi^+\pi^-$, we can get a lower bound
for $-1 \le \cos \theta \le -a$, $0 \le a < 1$.
Spin complications can be overcome. Following Mahoux and Martin \cite{mahmar68}
we can take as amplitude the sum of all diagonal helicity amplitudes
which has both in the $s$ channel and the $u$ channel the right positivity
properties of the absorptive parts.


\appendix

\section{Appendix. Lower bounds for the particle-particle and
particle-antiparticle forward scattering amplitudes}
\label{lowerbd}

In \cite{jinmartin64} a lower bound for a crossing-symmetric forward
scattering amplitude was obtained. This is the case for, for instance,
the $\pi_0\ p \rightarrow \pi_0\ p$ scattering amplitude.
If the scattering amplitude is not crossing-symmetric,
we can always symmetrize it, but then one gets  a lower bound {\it only}
on the average, say $\haf [AB \rightarrow AB + A\bar B \rightarrow A\bar B]$,
and we can only say that it applies to one of the amplitudes, but one
does not know which one. Here, at the price of a very small weakening
of the lower bound, we get lower bounds separately for $AB \rightarrow AB$
and $A\bar B \rightarrow A\bar B$.

\vskip 3mm
We assume that the forward scattering amplitude satisfies a
dispersion relation with a finite number of subtractions $N$.
Above the right-hand cut $\Im F_{AB \rightarrow AB}(s) >0$.
On the left-hand cut $\Im F_{A\bar B \rightarrow A\bar B}(u) >0$,
with $u = 2(M_A^2 + M_B^2) -s$, which means that, {\it above} the 
left-hand cut, the imaginary part is negative.

\begin{figure}[ht]
\begin{center}
\includegraphics{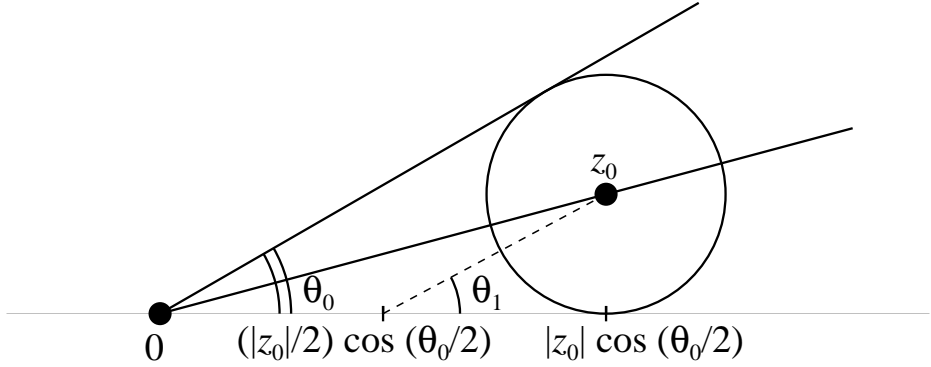}
\caption {The point $z_0$}
\label{figapx1}
\end{center}
\end{figure}

First we study a function $G(z) = \ovl{G(\bar z)}$
with a {\it positive} imaginary part
above both cuts, with a finite number of subtractions $N$. We may suppose
$N$ even (otherwise we use $N+1$). $\Im G(z)$ vanishes in an open interval
containing 0.
\begin{align}
G(z) &= \sum_{n=0}^{N-1} c_n z^n 
+ {z^N\over \pi} \int_{-\infty}^{+\infty} {\Im G(z')\,dz'\over
(z'-z){z'}^N}\ ,\cr
&\Im G(z) \ge 0\ \ {\rm for\ \ real}\ z\ .
\label{x.10}\end{align}
First we shall prove that 
${G(iy)\over y^N} \rightarrow 0$ when $y \rightarrow +\infty$.
\beq
G(iy) = \sum_{n=0}^{N-1} c_n (iy)^n
+ {(iy)^N\over \pi} \int_{-\infty}^{+\infty} {\Im G(z')\,dz'\over
(z'- iy){z'}^N}\ .
\label{x.15}\endq
\begin{align}
|G(iy)| &\le \ \hbox{Polynomial of degree $N-1$} \cr
&+ {y^N\over \pi} \int_{-\infty}^{-M} {\Im G(z')\,dz'\over
|z'|^{N+1}} \cr
&+ {y^N\over \pi} \int_{M}^{\infty} {\Im G(z')\,dz'\over
z'^{N+1}} \cr
&+ {y^{N-1}\over \pi} \int_{-M}^M {\Im G(z')\,dz'\over
|z'|^{N}}\ .
\label{x.16}\end{align}
By taking $M$ large enough, we can make the first two terms
less than $\veps y^N$. The third term is bounded by
$Cy^{N-1} = {1\over y}\,Cy^N$. So, for $y \rightarrow \infty$,
$\left | {G(iy)\over y^N}\right | \le 3\veps$, but $\veps$ can be taken
arbitrarily small, so
\beq
\lim_{y\rightarrow \infty}\left | {G(iy)\over y^N}\right | = 0\ .
\label{x.17}\endq
Suppose we have at the same time
\beq
\lim \left | {G(z)\over z^N} \right | \rightarrow 0\ \ \ 
{\rm for}\ \ |z| \rightarrow  \infty,\ \ \Arg z ={\pi\over 2}\ ,
\label{x.20}\endq
and
\beq
G(z)|z|^{1+\alpha} \rightarrow 0\ \ \ 
{\rm for}\ \ |z| \rightarrow  \infty,\ \ \Arg z = 0\ .
\label{x.21}\endq
Then construct $H$:
\beq
H(z) = G(z)\exp \left [ (1+\alpha)\log z
+{i\over \pi}(N+1+\alpha)(\log z)^2 \right ]\ .
\label{x.25}\endq
$H(z)$ tends to 0 as $|z|\rightarrow  \infty$ both for $\Arg z =0$
and $\Arg z = {\pi\over 2}$. By the Phragm\'en-Lindel\"of Theorem
$H(z)$ tends to 0 as $|z|\rightarrow  \infty$ for
$0\le \Arg z \le {\pi\over 2}$. So
\beq
\lim_{|z|\rightarrow  \infty}
|G(z)| |z|^{(1+\alpha -{2\over \pi}\theta (1+\alpha+ N))} = 0,\ \ \ 
\theta = \Arg z \in \left [0,\ {\pi\over 2}\right ]\ .
\label{x.26}\endq
In particular
\beq
\lim_{|z|\rightarrow  \infty}|G(z)|| z|^{\left (1+{\alpha\over 2}\right )} = 0\ \ \ 
{\rm for}\ \ 0\le \theta \le \theta_0 = {\pi\over 4}
{\alpha \over 1+\alpha+N}\ .
\label{x.30}\endq
In what follows we shall eliminate subtractions by differentiating $G$
$N+1$ times. We need a bound on $\left({d\over dz}\right )^{N+1}G(z)$
somewhere in the angle $0 < \theta < \theta_0$. We shall take
$z_0 = |z_0|e^{i{\theta_0\over 2}}$. The disk $|z-z_0| < |z_0|\sin(\theta_0/2)$
is contained in the angular interval $0 < \theta < \theta_0$
(see Fig.~\ref{figapx1}).
\vglue 1cm

Using the Cauchy integral for $\left ({d\over dz}\right )^{N+1}G(z)$
at $z= z_0$ we get
\beq
\lim_{z_0 \rightarrow \infty}\left [ |z_0| \left (1-\sin {\theta_0\over 2}
\right ) \right ]^{\left ( 1+{\alpha\over 2} \right )}
{\left | z_0 \sin\left ({\theta_0\over 2} \right ) \right |^{N+1}\over
(N+1)!}\left ({d\over dz}\right)^{N+1}G(z_0) = 0\ . 
\label{x.35}\endq
However we can also estimate $\left ({d\over dz}\right )^{N+1}G(z)$
from the dispersion relation
\beq
\left ({d\over dz}\right)^{N+1}G(z) = {(N+1)!\over\pi}\int_{-\infty}^\infty
{\Im G(z')\,dz' \over (z'-z)^{N+2}}\ . 
\label{x.40}\endq
We shall find a lower bound for the real part of this quantity.
Here the positivity of $\Im G(z')$ in the integral is essential,
and the evenness of $N$ is used.

We suppose $0< \alpha < 1$.
If $z' \le \haf |z_0|\cos {\theta_0\over 2}$ we have (see Fig. \ref{figapx1})
\beq
0 < \Arg(z_0-z') \le \theta_1 = \Arg\left (z_0-\haf |z_0|\cos {\theta_0\over 2}
\right ) < \theta_0 = {\pi\over 4} {\alpha \over 1+\alpha+N}\ ,
\label{x.41}\endq
\beq
0 < \Arg\left ((z_0-z')^{N+2} \right ) < 
{\pi\over 4} {(N+2)\alpha \over 1+\alpha+N} < {\alpha\pi\over 2}\ .
\label{x.42}\endq
Therefore
\begin{align}
\Re \left ({d\over dz}\right)^{N+1}G(z_0) &\ge {(N+1)!\over\pi}
\int_{-\infty}^{\haf |z_0|\cos {\theta_0\over 2}}
{\Im G(z')\, \cos (\alpha\pi/2)\, dz' \over
|z'-z_0|^{N+2}} \cr
& - {(N+1)!\over\pi} \int_{\haf |z_0|\cos {\theta_0\over 2}}^\infty
{1\over \left |z_0 \sin{\theta_0\over 2} \right |^{N+2}}
\Im G(z')\, dz' \ ,
\label{x.45}\end{align}
but $\Im G(z') < C|z'|^{-(1+\alpha)}$ from (\ref{x.21}), so that the
second term is less than $|z_0|^{-(N+2+\alpha)}$. The first term 
is larger than
\beq 
{C\over |z_0+M|^{N+2}} \int_{-M}^M \Im G(z')\,dz'\ , 
\label{x.46}\endq
so that, for $|z_0|$ large, the second term is negligible, 
but then (\ref{x.46}) 
contradicts (\ref{x.35}) . 
The conclusion is that 
\beq
\limsup_{z \rightarrow \infty} |G(z)|z^{1+\alpha} > 0,\ \ \alpha > 0\ \ 
\hbox{arbitrarily small}\ ,
\label{x.50}\endq
in fact it reaches infinity.

\subsection{Application to the scattering amplitude}
$F(s)$ satisfies a dispersion relation with two cuts corresponding
to the processes $AB\rightarrow AB$ and $A\bar B\rightarrow A\bar B$.
Define $z = s - M_A^2 - M_B^2$. We keep the notation $F(z)$.
Since the imaginary part above the right-hand cut is positive
and the imaginary part {\it above} the left-hand cut is negative,
multiply $F(z)$ by $z$ :
\beq
G(z) = zF(z)\ .
\label{x.60}\endq
Then the previous results apply and we have {\it separately}
\beq
\limsup_{s\rightarrow +\infty} |F(s)|\,s^{2+\alpha} > 0\ ,
\label{x.61}\endq
$\alpha > 0$ arbitrarily small, for the reaction $AB\rightarrow AB$
and
\beq
\limsup_{u\rightarrow +\infty} |F(u)|\,u^{2+\alpha} > 0\ ,
\label{x.62}\endq
for the reaction $A\bar B\rightarrow A\bar B$.

Let us notice that these results persist if we convolute $F$ with a
positive function with compact support $w$ :
\beq
F_w(s) = \int F(s') w(s-s')\,ds'\ .
\label{x.65}\endq


\section{Appendix. A problem of estimation}
\label{interp}

Let $f$ be a function holomorphic on the domain $D_L$ bounded by the ellipse
$E_L$ with foci $\pm 1$ and semi-great axis $\ch(L)$, $L>0$,
\begin{align}
E_L &= \{ z\ :\ z = \cos(\theta + iL),\ \theta \in \bR\}\ ,\cr
D_L &= \{ z\ :\ z = \cos(\theta + iy),\ \theta \in \bR,\ \ 
y \in \bR,\ \ |y| < L\}\ .
\label{a.10}\end{align}
We suppose
\beq
|f(z)| \le M\ \ \hbox{for all}\ \ z \in D_L\ ,\ \
|f(z)| \le m \ \ \hbox{for all}\ \ z \in  [-a,\ a]\ ,
\label{a.20}\endq
where $0< m < M$ and $a = \cos(b)$, $0< a< 1$, $0<b< \pi/2$. 
We seek an upper bound for $|f(1)|$. 
General theorems (see e.g. \cite[pp 141-145]{B}) assert the existence and
uniqueness of a function $H$, harmonic on $D_L\setminus [-a,\ a]$
(i.e. $D_L$ minus a cut along the segment $[-a,\ a]$) and continuous
at the boundary, and such that $H=1$ on $E_L$ and $H=0$ on $[-a,\ a]$.
For every $z \in D_L\setminus [-a,\ a]$, $0< H(z) = H(-z) = H(\bar z) < 1$ and
\beq
\log |f(z)| \le H(z)\log(M) + [1-H(z)]\log(m)\ \ \ \forall z \in D_L\ .
\label{a.25}\endq
Indeed $\log |f(z)|$ is subharmonic in $D_L\setminus [-a,\ a]$, and
at the boundary it is majorized by the harmonic function which appears
on the rhs of (\ref{a.25})(See e.g. \cite[pp 16-18]{H}, \cite[p. 132]{B}).
In particular
\beq
\log |f(1)| \le H(1)\log(M) + [1-H(1)]\log(m)\ .
\label{a.26}\endq
It is possible to give an exact determination of $H$
but we will give cruder upper and lower bounds for it which
describe more explicitly its behavior when $L$ tends to 0.
We use the notation:
\beq
\bC_+ = - \bC_- = \{z \in \bC\ :\ \Im z >0\},\ \ \ \
{\bf D} = \{z \in \bC\ :\ |z|< 1\}\ .
\label{a.26.1}\endq
By a conformal map we always mean a holomorphic injective map.

\subsection{Upper and lower bounds for $H$}
\label{m1}
Let $h(z) = H(\cos(z))$. This function is harmonic, even,
and has period $\pi$ in the domain
\begin{align}
& S_L \setminus \bigcup_{n\in\bZ} (I+n\pi)
\ ,
\label{a.27.1}\\
&S_L = \{z\ :\ |\Im z| < L\}\ ,
\label{a.27.2}\\
& I = [b,\ b']\ ,\ \ b'= \pi-b > {\pi\over 2} > b\ .
\label{a.27.3}\end{align}
Note that $-I = I-\pi$,
and that
$\bigcup_{n\in\bZ} (I+n\pi) = \cos^{-1}([-a,\ a])$.
In the domain (\ref{a.27.1}), $0 < h(z)<1$.
$h$ is continuous at the boundary of this domain,
and takes the value 1 on the edges of $S_L$, i.e. $\bR \pm iL$ and the value
0 on the cuts $I+n\pi$. 
Note that $H(1) = h(0)$.

\begin{figure}[ht]
\begin{center}
\includegraphics{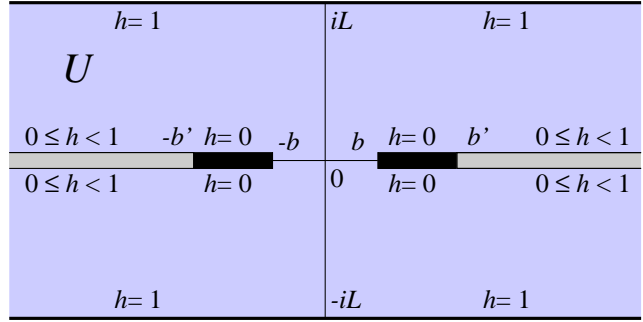}
\caption {The domain $U$ and boundary values for $h$}
\label{fig1}
\end{center}
\end{figure}

Let $U$ be the smaller domain consisting
of the strip $S_L$ minus two cuts on $(-\infty,\ -b]$ and $[b,\ +\infty)$.
$h$ is harmonic in $U$ and continuous at its boundary (See Fig. \ref{fig1}). 

We can conformally map
$U$ onto the upper half-plane by a map $\psi_2\circ\psi_1$.

\begin{figure}[ht]
\begin{center}
\includegraphics{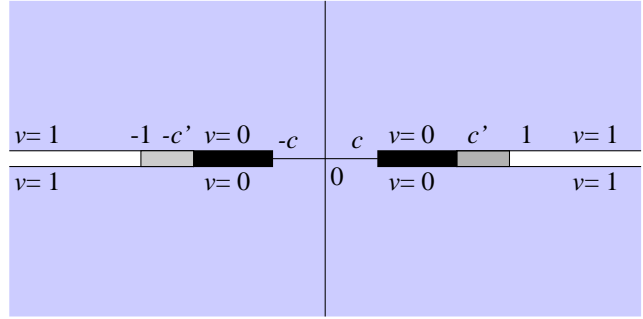}
\caption {Domain and boundary values for $v$}
\label{fig2}
\end{center}
\end{figure}

\vskip 3mm
The map $\psi_1$ conformally maps the strip $S_L= \{z\ :\ |\Im z| < L\}$
onto the cut-plane $\bC_+\cup\bC_-\cup (-1,\ 1)$ :
\beq
Z=\psi_1(z)\ \ \ \Longleftrightarrow
\ \ z= {L\over\pi} \log\left({1+Z\over 1-Z}\right ),\ \ \
Z = \th \left ( {\pi z\over 2L}\right )\ .
\label{e.40}\endq
It maps the cut-strip $U$ pictured in Figure \ref{fig1} onto the cut-plane
$\bC_+\cup \bC_-\cup (-c,\ c)$, where
\beq
c = \psi_1(b) = \th \left ({\pi b\over 2L}\right )\ ,\ \ \ \
c' = \psi_1(b') = \th \left ({\pi b'\over 2L}\right )\ .
\label{e.41}\endq
Recall that $b' = \pi-b > b$ so that $1> c' > c$. Denoting
$v(Z) = h(\psi_1^{-1}(Z))$, the domain and boundary values for
$v$ are pictured in Fig. \ref{fig2}. The points $Z= \pm 1$
are images of $z = \pm\infty$ so that $v$ is not continuous there.
It is continuous at all the other boundary points of the cut-plane.

The map $\psi_2$ conformally maps the cut-plane $\bC_+\cup \bC_-\cup (-c,\ c)$
(pictured in Fig. \ref{fig2}) onto the upper half-plane $\bC_+$.
\beq
\zeta=\psi_2(Z)\ \ \ \Longleftrightarrow\ \ \ 
Z = {c\over 2}\left ( \zeta + {1\over \zeta}\right ),\ \ \ 
\zeta = {Z\over c} + \sqrt {\left ({Z\over c}\right )^2 -1}\ .
\label{e.50}\endq
In this formula the function $t \mapsto \sqrt{t^2-1}$ is defined to be
holomorphic
with a positive imaginary part in the cut-plane $\bC_+\cup\bC_-\cup (-1,\ 1)$.
Hence $\psi_2(\bar Z) = 1/\ovl{\psi_2(Z)}$.
If $t$ is real with $t>1$, then
\begin{align}
&(t+i0) + \sqrt{(t+i0)^2-1} = t+|\sqrt{t^2-1}| +i0,
\label{e.55}\\
&(t-i0) + \sqrt{(t-i0)^2-1} = t-|\sqrt{t^2-1}| +i0
= {1\over t+|\sqrt{t^2-1}|} + i0.
\label{e.56}\end{align}
The image $\psi_2(\bC_-)$ of the lower half-plane is ${\bf D}\cap \bC_+$,
$\psi_2(0) = i$, and $\psi_2(-i\infty) = 0$.
Let $A_0= \psi_2(1-i0)$, $B_0=  \psi_2(1+i0)$, $A= \psi_2(c'-i0)$,
$B=  \psi_2(c'+i0)$, i.e.
\begin{align}
&B_0 = {1\over c} + \sqrt{\left ({1\over c} \right )^2 -1} =
\ {1\over A_0}\ ,\ \ \
B = {c'\over c} + \sqrt{\left ({c'\over c} \right )^2 -1} =
\ {1\over A}\ ,\cr
&A_0 < A < 1 < B < B_0\ .
\label{a.40}\end{align}

Let
\beq
u(\zeta) = v(\psi_2^{-1}(\zeta)) =
h(\psi_1^{-1}(\psi_2^{-1}(\zeta)))\ .
\label{e.57}\endq
In other words $u$ is the result of transporting $h$ by the successive
coordinate changes $z \rightarrow Z \rightarrow \zeta$. In particular
$u(i) = h(0) = H(1)$. 
$u$ is harmonic in the upper half-plane
$\bC_+$ and $0< u(\zeta) < 1$ for all $\zeta \in\bC_+$. The function
$1-u$ has the same properties. $1-u$ is continuous at the real points
except at 0 and $\pm A_0$ and $\pm B_0$. Its boundary values at other
points are :
\begin{align}
& 1-u(t+i0) = 1\ \ \ {\rm for}\ \ t\in [-B,\ -A]\cup [A,\ B]\ ,\cr
& 1-u(t+i0) = 0\ \ \ {\rm for}\ \ t\in \bR \setminus \Big ([-B_0,\ -A_0]\cup
[A_0,\ B_0] \cup \{0\} \Big )\ ,\cr
& 0< 1-u(t+i0) \le 1\ \ {\rm for}\ \ t\in (-B_0,\ -B)\cup (-A,\ -A_0)
\cup (A_0,\ A) \cup (B,\ B_0)\ .
\label{e.60}\end{align}

\vskip 3mm
If $x_1$ and $x_2$ are real with $x_1< x_2$, let
\beq
\chi_{x_1,x_2} (\zeta) = \log\left ({\zeta-x_2\over \zeta-x_1}\right )
\label{e.70}\endq
be defined as holomorphic in $\bC \setminus [x_1,\ x_2]$,
and mapping $\bC_+$ (resp. $\bC_-$) into itself.
If $\zeta \in \bC_+$, $\Im \chi_{x_1,x_2} (\zeta)$ is in $(0,\ \pi)$
and is the angle under which the segment $(x_1,\ x_2)$ is seen
from the point $\zeta$. It is a harmonic function in $\bC_+$,
continuous at all real points except $x_1$ and $x_2$,
with boundary values equal to 0 outside of $[x_1,\ x_2]$,
and to $\pi$ on $(x_1,\ x_2)$. It tends to 0 at infinity in the closed
upper half-plane.

For $\zeta\in\bC_+$, let $u_+(\zeta)$ and $u_-(\zeta)$ be defined by
\beq
1-u_+(\zeta) = {1\over \pi}\Im \chi_{-B,-A}(\zeta)
+{1\over \pi}\Im \chi_{A,B}(\zeta)\ ,
\label{e.75}\endq
\beq
1-u_-(\zeta) = {1\over \pi}\Im \chi_{-B_0,-A_0}(\zeta)
+{1\over \pi}\Im \chi_{A_0,B_0}(\zeta)\ .
\label{e.76}\endq
For $\zeta\in \bC_+$, $1-u_\pm(\zeta) \in (0,\ 1)$ since the angles under
which $[-B_0,\ -A_0]$ and $[A_0,\ B_0]$ are seen from $\zeta$ add up to less
than $\pi$.

\begin{figure}[ht]
\begin{center}
\includegraphics{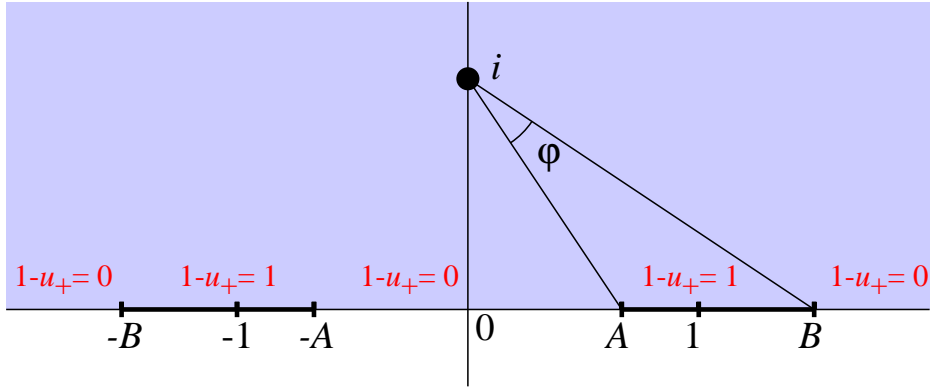}
\caption {Domain and boundary values for $1-u_+$. $\vhi = \Im\chi_{A,B}(i)$.
The picture for $1-u_-$ is the same with $A_0$ and $B_0$ instead of
$A$ and $B$}
\label{fig3}
\end{center}
\end{figure}

In particular, recalling that $A=1/B$,
\beq
\Im \chi_{-B,-A}(i) = \Im \chi_{A,B}(i) = 
\Arctg(B) - \Arctg(A) = 2\Arctg(B)- {\pi\over 2}\ ,
\label{e.77}\endq
\beq
1-u_+(i) = {4\over\pi}\Arctg(B)-1 \ .
\label{e.78}\endq
See Fig. \ref{fig3}. Similarly
\beq
1-u_-(i) = {4\over\pi}\Arctg(B_0)-1 \ .
\label{e.79}\endq
The boundary values of $1-u_+$ are (see Fig. \ref{fig3}) :
\beq
1-u_+(t+i0) = 1\ \ \ {\rm for}\ \ t\in (-B,\ -A)\cup (A,\ B)\ ,
\label{e.62}\endq
\beq
1-u_+(t+i0) = 0\ \ \ {\rm for}\ \ t\not\in [-B,\ -A]\cup [A,\ B]\ ;
\label{e.62.1}\endq
Those of $1-u_-$ are :
\beq
1-u_-(t+i0) = 1\ \ \ {\rm for}\ \ t\in (-B_0,\ -A_0)\cup (A_0,\ B_0)\ ,
\label{e.63}\endq
\beq
1-u_-(t+i0) = 0\ \ \ {\rm for}\ \ t\not\in [-B_0,\ -A_0]\cup [A_0,\ B_0]\ ;
\label{e.64}\endq
Thus, except at a finite set of real points, we have
\beq
1-u_+(\zeta+i0) \le 1-u(\zeta+i0) \le 1-u_-(\zeta+i0),\ \ \zeta \in \bR\ .
\label{e.65}\endq
In spite of the exceptional points
we can still apply the maximum (or minimum) principle in the form of
the following lemma 

\begin{lemma}
\label{max}
Let $g$ be a continuous function on the closed upper half-plane
with the exception of a finite set $\FF$ of real points.
We suppose that $|g(\zeta)| < C < \infty$ and $g$ is harmonic in $\bC_+$,
and $g(\zeta) \ge 0$ for all $\zeta \in \bR\setminus \FF$. Then
$g(\zeta) \ge 0$ for all $\zeta\in\bC_+$. If we assume that
$g(\zeta) = 0$ for all $\zeta \in \bR\setminus \FF$, then
$g(\zeta) = 0$ for all $\zeta \in \bC_+$.
\end{lemma}

(The boundedness of $|g|$ is essential as the example of
$\zeta\mapsto -\Im \zeta$ shows).
To prove this lemma, choose a fixed $\zeta_0 \in \bC_+$. The function
$\vhi(\zeta) = i(\zeta - \zeta_0)/(\zeta - \bar\zeta_0)$ maps $\bC_+$
onto ${\bf D}$, the closed upper half-plane onto
$\ovl{\bf D}\setminus \{i\}$, and $\vhi(\zeta_0) = 0$. The function
$G(z) = g(\vhi^{-1}(z))$ is harmonic in ${\bf D}$ and
continuous on $\ovl{\bf D}\setminus  \FF_1$ where $\FF_1$ is a finite
subset of the unit circle. Wherever defined, $|G(z)| \le C$. 
Let $\FF_2 = \{\theta\in [0,\ 2\pi]\ :\ e^{i\theta} \in \FF_1\}$.
For sufficiently small $\k\in(0,\ 1)$ there is a compact subset $E_{\k}$
of $[0,\ 2\pi]$ whose complement contains $\FF_2$ and has measure $\le 2\pi\k$,
and an $r_{\k} \in (0,\ 1)$ such that
$|G(e^{i\theta}) - G(r_{\k}e^{i\theta})|< \k$ for all $\theta \in E_{\k}$.
For $\theta\in E_{\k}$, $G(e^{i\theta}) \ge 0$. Therefore
\begin{align}
  G(0) &= \int_0^{2\pi}G(r_{\k}e^{i\theta}){d\theta\over 2\pi}\cr
&\ge 
\int_{E_{\k}} G(r_{\k}e^{i\theta}){d\theta\over 2\pi} - C\k \ge
\int_{E_{\k}} G(e^{i\theta}){d\theta\over 2\pi} - \k -C\k \ge -(C+1)\k\ .
\label{l.10}\end{align}
Letting $\k$ tend to 0 we get $G(0) = g(\zeta_0) \ge 0$. If we assume that
$g(\zeta) = 0$ for all $\zeta \in \bR\setminus \FF$, then also
$-g(\zeta) \ge 0$ for all $\zeta \in \bC_+$, hence 
$g(\zeta) = 0$ for all $\zeta \in  \bC_+$.

\vskip 3mm
Applying this to $g = u_+-u$ and to $g= u-u_-$, we find
that $u_-(\zeta) \le u(\zeta) \le u_+(\zeta)$ for all $\zeta \in \bC_+$.
In particular $1-u(\zeta)$ tends to 0 if $\zeta$ tends to 0 or infinity
in the closed upper half-plane.
In fact the maximum principle implies that the inequalities are strict, i.e.
\beq
u_-(\zeta) < u(\zeta) < u_+(\zeta)\ \ \ \forall \zeta \in \bC_+\ .
\label{e.80}\endq
Hence
\beq
u_-(i) = 2- {4\over\pi}\Arctg(B_0) < u(i)= H(1) <
u_+(i) = 2- {4\over\pi}\Arctg(B)\ .
\label{e.81}\endq

\vskip 3mm
To study the behavior of $u_\pm(i)$ as $L\rightarrow 0$ we recall that,
for real $z \ge 0$,
\beq
0 \le {d\over dz} \Arctg(1+z) \le \haf,\ \ \
-{z^2\over 4} \le \Arctg(1+z) -{\pi\over 4} -{z\over 2} \le 0\ .
\label{an.10}\endq
We denote
\beq
x = \exp\left({-\pi b\over L}\right ),\ \ \ \
x'=  \exp\left({-\pi b'\over L}\right ),\ \ b' = \pi- b > b,\ \ x'<x\ .
\label{an.15}\endq
Note that $x'/x \rightarrow 0$ as $L\rightarrow 0$. With this notation
\begin{align}
&{1\over c} = {1+x\over 1-x},\ \ c' = {1-x'\over 1+x'}\ ,\ \ \ 
\sqrt{\left ( {1\over c}\right )^2 - 1} = {2\sqrt{x}\over(1-x)}\ ,\cr
& B_0 = {1\over c} +\sqrt{\left ({1\over c}\right )^2 -1}
= {1+\sqrt{x}\over 1-\sqrt{x}} = 1 +{2\sqrt{x}\over 1-\sqrt{x}}\ .
\label{an.20}\end{align}
Applying (\ref{an.10}) with $z = 2t/(1-t)$, $t = \sqrt{x}$ gives
\begin{align}
& -{t^3\over (1-t)^2}\le \Arctg(B_0) - {\pi\over 4} - t \le {t^2\over 1-t}\ ,\cr
& \left |\Arctg(B_0) - {\pi\over 4} - t\right | \le
{t^2\over (1-t)^2}\ .
\label{an.25}\end{align}
Hence
\beq
\left |u_-(i) - 1 +{4\sqrt{x}\over\pi}\right |
\le {4x \over \pi(1-\sqrt{x})^2}\ ,\ \ x = \exp\left({-\pi b\over L}\right )\ .
\endq
Further
\begin{align}
&{1\over c} - {c'\over c} = {2x'\over c(1+x')} < {2x'\over c}\ ,\ \ \ 
\left ( {1\over c}\right )^2 - \left ({c'\over c} \right )^2 =
{4x'\over c^2(1+x')^2} < {4x'\over c^2}\ ,\cr
& \sqrt{\left ( {1\over c}\right )^2 - 1}
- \sqrt{\left ( {c'\over c}\right )^2 - 1} < {{4x'\over c^2}\over
\sqrt{\left ( {1\over c}\right )^2 - 1}}
= {2x'(1+x)\over c\sqrt{x}}\ ,\cr
&B_0-B < {2x'\over c}\left [ 1+ {1+x\over \sqrt{x}} \right ]<
2\sqrt{x'}\left [ 1+ \sqrt{x} +x \right ] 
\left ({1+x\over 1-x} \right )\ ,\cr
&\Arctg(B_0) - \Arctg(B) < \sqrt{x'}\left [ 1+ \sqrt{x} +x \right ] 
\left ({1+x\over 1-x} \right )\ .
\label{an.30}\end{align}
Hence
\begin{align}
H(1) - u_-(i) \le
u_+(i) - u_-(i) &< {4\sqrt{x'}\over\pi}\left [ 1+ \sqrt{x} +x \right ] 
\left ({1+x\over 1-x} \right )\cr
&= {4\sqrt{x'}\over\pi}(1+O(\sqrt{x}))\ .
\label{an.35}\end{align}
This gives
\beq
u_\pm(i) = 1- {4\over\pi} \exp\left({-\pi b\over 2L}\right)
+{\rm o}\left (\exp\left({-\pi b\over 2L}\right)\right)\ \ \
(L \rightarrow 0)\ ,
\label{a.80}\endq
\beq
H(1) = 1- {4\over\pi} \exp\left({-\pi b\over 2L}\right)
+{\rm o}\left (\exp\left({-\pi b\over 2L}\right)\right)\ \ \
(L \rightarrow 0)\ .
\label{a.81}\endq
Bounds on the error terms are supplied by the preceding inequalities.
Thus although some information is lost if $H(1)$ is replaced by
its upper bound $u_+(i)$, this becomes unimportant for very small $L$.

We also note that if we define $h_\pm(z) = u_\pm(\psi_2(\psi_1(z)))$,
\beq
h_-(z) < h(z) < h_+(z)\ \ \forall z\in U\ .
\label{a.82}\endq

\begin{figure}[ht]
\begin{center}
\includegraphics{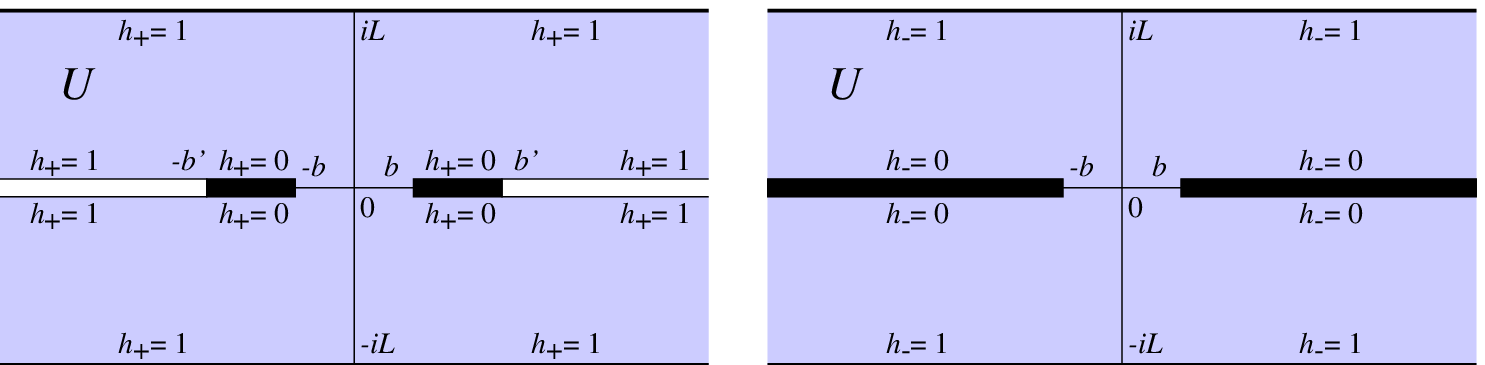}
\caption {Domain and boundary values for $h_+$ (left) and $h_-$ (right)}
\label{fig11}
\end{center}
\end{figure}

The bounds obtained in this section are not useful at large $L$. In fact
when $L\rightarrow \infty$, $(c'/c)\rightarrow (b'/b)$ so that
$u_+(i)$ tends to a non-zero limit, while it can be shown that
$H(1)$ tends to 0

\end{document}